**Excitation of self-localized spin-wave "bullets" by spin-polarized current in in-plane magnetized magnetic nano-contacts: a micromagnetic study**

G. Consolo[a*], B. Azzerboni[a], G. Gerhart[b], G.A. Melkov[c], V. Tiberkevich[d], and A.N. Slavin[d]

[a] *University of Messina, Dipartimento di Fisica della Materia e Tecnologie Fisiche Avanzate, Salita Sperone 31, Vill.S.Agata, Messina 98166, ITALY*

[b] *U.S. Army TARDEC, Warren, Michigan 48397, USA*

[c] *National Taras Shevchenko University of Kiev, Faculty of Radiophysics, Kiev 01033, Ukraine*

[d] *Oakland University, Departament of Physics, Rochester, Michigan 48309, USA*

(Dated: 25 May 2007)

Abstract

It was shown by micromagnetic simulation that a current-driven in-plane magnetized magnetic nano-contact, besides a quasi-linear propagating ("Slonczewski") spin wave mode, can also support a nonlinear self-localized spin wave "bullet" mode that exists in a much wider range of bias currents. The frequency of the "bullet" mode lies below the spectrum of linear propagating spin waves, which makes this mode evanescent and determines its spatial localization. The threshold current for the excitation of the self-localized "bullet" is substantially lower than for the linear propagating mode, but finite-amplitude initial perturbations of magnetization are necessary to generate a "bullet" in our numerical simulations, where thermal fluctuations are neglected. Consequently, in these simulations the hysteretic switching between the propagating and localized spin wave modes is found when the bias current is varied.

# INTRODUCTION

It was theoretically predicted[1-3] and experimentally observed[4-9] that persistent microwave magnetization precession can be excited in a thin ("free") layer of a magnetic layered structure by direct current traversing the structure. The bias current passing through a magnetic layered structure becomes spin-polarized in the direction of magnetization of a thicker ("fixed") magnetic layer, and then can transfer this induced spin angular momentum to the magnetization of a thinner ("free") magnetic layer. For the proper direction of the bias current this spin-transfer mechanism creates an effective negative magnetic damping in the "free" magnetic layer, which, for sufficiently large current magnitude, can compensate the natural positive magnetic damping and lead to the excitation of microwave spin waves[3,10,11].

The analytical theory of spin wave excitation in magnetic nano-contacts by spin-polarized current performed in linear[3] and weakly nonlinear[12] approximations showed possibility of self-sustained excitation of two qualitatively different modes: linear propagating "Slonczewski" mode[3] and nonlinear evanescent "bullet" mode[12]. The latter mode exists only in in-plane magnetized case, has a substantially lower excitation threshold due to its self-localized character and, consequently, vanishing radiation losses, and is believed[12] to have been observed in experiments[7-9].

At the same time, the full-scale micromagnetic simulations of magnetization dynamics in in-plane magnetized nano-contacts[13,14] done using the Landau-Lifshitz-Gilbert equation with Slonczewski spin-transfer term showed no self-sustained excited spin wave states for current densities *below* the threshold of excitation of a linear propagating "Slonczewski" spin wave mode.

Thus, it still remains unclear whether the analytically predicted low-threshold bullet mode[12] is an artifact of the small-amplitude expansion of full equations of motion for the magnetization done in Ref. 12, or it is a physical reality that can be observed in in-plane magnetized nano-contacts. In the latter case it is necessary to understand why the micromagnetic simulations[13,14] failed to reproduce this low-threshold localized spin wave mode.

It should be noted that a spin wave mode having properties similar to the properties of a self-localized spin wave "bullet"[12] was found in numerical simulations[14], but for the current densities that were substantially larger than the instability threshold for the linear "Slonczewski" mode[3]. This high-current spin wave mode has many attributes of the self-localized nonlinear "bullet" mode: large precession angle, strong spatial localization, and low frequency (below the ferromagnetic resonance (FMR) frequency of the "free" layer). However, from the numerical results presented in Ref. 14 it is not clear whether this mode is really the spin wave "bullet"[12] or a strongly nonlinear spin wave excitation of a qualitatively different type related to the formation of vortex-antivortex pairs in a current-driven magnetic nano-contact[13-14].

The aim of our present paper is to verify the predictions of the analytical theory[12] about the existence of a low-threshold spin wave "bullet" mode using the full-scale micromagnetic simulations of the Landau-Lifshits-Gilbert-Slonczewski (LLGS) equation.

In contrast with the previous numerical studies[13,14], where the simulations of spin wave dynamics for each value of the bias current were performed starting from the equilibrium initial magnetization state, in our current work we, at first, progressively increase the bias current from zero to sufficiently large above-threshold value, and then progressively decrease this current to zero value. Using this method, we were able to observe in our simulations subcritically-unstable[15] spin wave modes (i.e. modes which require finite amplitude of spin wave fluctuations to be excited) even in the absence of thermal noise fluctuations. Starting our simulations from a large magnitude of the bias current (which corresponds to a strongly nonlinear regime of magnetization oscillations), and gradually reducing the current magnitude, we demonstrated that the spin wave "bullet" mode[12] can, indeed, be supported by bias currents that are substantially lower than the threshold of excitation of the linear "Slonczewski" mode[3]. At the same time, we also demonstrated that the spin wave "bullet" excitation is strongly subcritical (see Ref. 15 for details on subcritical instabilities) and, therefore, it is not possible to observe it when the bias current is increased starting from the equilibrium initial conditions (in the absence of thermal noise) up to relatively large magnitudes of the bias current. Only at the current magnitudes substantially exceeding the threshold current of a propagating "Slonczewski" spin wave mode the localized "bullet" mode with the frequency lower than the FMR frequency of the "free' layer is excited, in full agreement with the results of previous simulations[13,14]. Thus, the co-existence of two spin wave modes (propagating "Slonczewski" mode and localized "bullet" mode) with different critical currents and different instability scenario (linear and sub-critical) leads to the hysteretic behavior of a magnetic nano-contact when the bias current passing through it is varied.

## FORMULATION OF THE PROBLEM

We studied current-induced spin wave dynamics in a magnetic multi-layered system consisting of a thick magnetic "pinned" layer (PL, see Fig. 1) that serves as a spin polarizer, a thin non-magnetic spacer, and a thin magnetic "free" layer (FL). The thickness of the PL is assumed to be large enough to prevent any dynamics in this layer. The bias magnetic field $H$ is applied in the plane of the structure along the axis $z$. The bias current $I$ traversing the multi-layered structure is applied within the circular nano-contact area of the radius $R_c$ (see Fig. 1).

The dynamics of magnetization $\mathbf{M} = \mathbf{M}(t, \mathbf{r})$ of the "free" magnetic layer under the action of spin-polarized current is described by the Landau-Lifshitz-Gilbert-Slonczewski (LLGS) equation:

$$\frac{\partial \mathbf{M}}{\partial t} = \gamma [\mathbf{H}_{\text{eff}} \times \mathbf{M}] + \frac{\alpha(\xi)}{M_0} \left[ \mathbf{M} \times \frac{\partial \mathbf{M}}{\partial t} \right] + f(r/R_c) \frac{\sigma I}{M_0} [\mathbf{M} \times [\mathbf{M} \times \mathbf{p}]], \tag{1}$$

where $\gamma$ is the gyromagnetic ratio and $\mathbf{H}_{\text{eff}}$ is the effective magnetic field calculated as a variational derivative ($\mathbf{H}_{\text{eff}}(t,\mathbf{r}) = -\delta W/\delta \mathbf{M}$) of the magnetic energy $W$ of the system, which includes magnetostatic, exchange, and Zeeman contributions.

The second term in the right-hand side of Eq. (1) is the phenomenological magnetic damping torque written in the nonlinear form[16] that is similar, but not identical, to the traditional Gilbert form used in the previous simulations[13,14], and $M_0 = |\mathbf{M}|$ is the saturation magnetization of the "free" layer. Since the spin-torque mechanism of spin wave excitation is very efficient, it can lead to rather large magnetization precession angles very soon above the excitation threshold (see e.g. Fig. 10 in Ref. 10). For large precession angles the Gilbert damping parameter can not be considered constant anymore, and should be replaced by the "damping function" $\alpha(\xi)$ having the form[16]

$$\alpha(\xi) = \alpha_G (1 + q_1 \xi), \qquad (2)$$

where $\alpha_G$ is the dimensionless Gilbert damping constant, $q_1$ is a dimensionless phenomenological nonlinear damping parameter of the order of unity, and $\xi$ is the dimensionless variable characterizing level of nonlinearity of magnetization precession (see Ref. 16 for details):

$$\xi = \frac{(\partial \mathbf{M}/\partial t)^2}{\omega_M^2 M_0^2}, \qquad (3)$$

where $\omega_M = \gamma M_0$.

To determine how important is the role of nonlinear damping in the current-induced spin wave excitation, we used in our simulations two different values of the parameter $q_1$: $q_1 = 0$, which gives us the classical Gilbert damping model and $q_1 = 3$, which corresponds to a moderate degree of damping nonlinearity.

The last term in the right-hand side of Eq. (1) is the Slonczhewski spin-transfer torque[1,3] that is proportional to the bias current $I$. The function $f(r/R_c)$ characterizes the distribution of current across the nano-contact area. In the simplest case of uniform current density distribution $f(r/R_c) = 1$ if $r < R_c$ and $f(r/R_c) = 0$ otherwise. In Eq. (1) the proportionality coefficient $\sigma$ is determined by the spin-polarization efficiency $\varepsilon$ and is given by the expression[3, 10]

$$\sigma = \frac{\varepsilon g \mu_B}{2 e M_0 S d}, \qquad (4)$$

where $\varepsilon$ the dimensionless spin-polarization efficiency defined in Refs. 1 and 3, $g$ is the Landè factor, $\mu_B$ is the Bohr magneton, $e$ is the absolute value of the electron charge, $d$ is the FL thickness and $S = \pi R_c^2$ is the cross-sectional area of the nano-contact. In Eq. (1) the unit vector **p** defining the spin-polarization direction is parallel to the direction $\mathbf{e}_z$ of the in-plane external magnetic field.

In our calculations we made several simplifying assumptions. First of all, we neglected the constant current-induced (Oersted) magnetic field and the magnetostatic coupling between the two ferromagnetic layers (FL and PL) of a nano-contact as we do not believe that in the presence of a sufficiently large constant bias magnetic field these effects can qualitatively change the structure of spin wave modes excited in a nano-contact by a spin-polarized current. Second, we assumed that the magnetocrystalline anisotropy of the "free" layer is negligibly small.

To reduce the computation time we, also, neglected the random fluctuations arising from the thermal noise. Our further investigations have shown that, although these fluctuations do not change the structure of spin wave modes that could be excited in a nano-contact, they might play an important role in the process of excitation of a particular spin wave mode in a laboratory experiment.

In our simulations we used a set of material parameters that is typical for the experiments with current-induced spin wave excitations in in-plane magnetized nano-contacts with permalloy free layer[8]: FL thickness $d = 5$ nm, nano-contact radius $R_c = 20$ nm, spin-polarization efficiency $\varepsilon = 0.25$, saturation magnetization of the FL $\mu_0 M_0 = 0.8$ T, external bias magnetic field $\mu_0 H = 0.5$ T, spectroscopic Landè factor $g = 2.0$, and FL exchange stiffness constant $A_{ex} = 1.4 \times 10^{-11}$ J/m. The

Gilbert damping constant was chosen to be $\alpha_G = 0.01$, and the nonlinear damping parameter $q_1$ was chosen to be $q_1 = 0$ in the standard Gilbert damping model and $q_1 = 3$ in the nonlinear damping model.

The *linear* analysis of current-induced spin wave excitations in the nano-contact geometry was performed in Ref. 3 for the case of perpendicular magnetization. This analysis was based on the *linearized* Eq. (1) and showed that a linear propagating spin wave mode ("Slonczewski" mode) is excited at the threshold. The threshold current for this propagating mode is determined by the sum of two contributions: radiation losses due to the propagation of the excited spin wave outside the region of current-carrying nano-contact, and dissipation of the current energy inside the nano-contact region [3]

$$I_{th}^L \approx 1.86 \frac{D}{\sigma R_c^2} + \frac{\Gamma}{\sigma}, \qquad (5)$$

where $D$ is the spin wave dispersion coefficient determined mostly by the exchange interaction and $\Gamma$ is the linear spin wave damping rate proportional to the Gilbert damping constant $\alpha_G$.

In the case of an *in-plane* magnetized magnetic FL linear propagating mode with the threshold current (5) can, also, exists, with parameters $\Gamma$ and $D$ having the form [12]: $\Gamma = \alpha_G(\omega_H + \omega_M/2)$ and $D = (2A_{ex}/M_0)(\omega_H + \omega_M/2)/\omega_{FMR}$, where $\omega_H = \gamma H$, $\omega_M = \gamma M_0$, $A_{ex}$ is the exchange constant, and $\omega_{FMR} = \sqrt{\omega_H(\omega_H + \omega_M)}$ is the ferromagnetic resonance (FMR) frequency in the "free" layer. For a typical nano-contact radius of the order of several tens of nanometers the main contribution to the linear threshold current (5) comes from the first term describing radiation losses. According to the linear theory[3], the propagating spin wave mode excited at the threshold is a cylindrical spin wave with the wave vector $k_L = 1.2/R_c$ and frequency

$$\omega_L = \omega_{FMR} + D\, k_L^2, \tag{6}$$

that is higher than the FMR frequency in the "free" layer. Due to its propagating character, the linear cylindrical spin wave mode excited at the threshold is relatively weakly localized near the excitation region (current-carrying nano-contact). Thus, in the limit of small damping the squared amplitude (proportional to the mode power) $A^2 = (M_0 - M_z)/2M_0$ of the linear "Slonczewski" mode decays with the radial distance $r$ as

$$A_L^2(r) \sim \frac{1}{r} \tag{7}$$

for $r \gg R_c$.

The *nonlinear* analysis[12] of spin wave excitations in *in-plane* magnetized nano-contact geometry revealed a qualitatively different picture. It was shown in Ref. 12 that the competition between the nonlinearity and exchange-related dispersion leads to the formation of a stationary two-dimensional self-localized non-propagating spin wave "bullet" mode whose frequency is shifted by the nonlinearity below the spectrum of linear spin wave modes, i.e. below the FMR frequency in the FL. This nonlinear mode has evanescent character with vanishing radiation losses, which leads to a substantial decrease of its threshold current in comparison to the linear propagating "Slonczewski" mode. In contrast with the linear mode, the "bullet" mode is strongly localized and its squared amplitude decays with the distance $r$ much faster then in the case of the linear propagating mode (7):

$$A_B^2(r) \sim \frac{1}{r} e^{-2|k_B|r}. \tag{8}$$

Here $k_B$ is an *imaginary* wave vector of the bullet mode, related to its frequency by the relation, similar to Eq. (6):

$$\omega_B = \omega_{FMR} - D|k_B|^2. \tag{9}$$

Although the analytical theory[12] based on the idea of spin wave "bullet" formation is in quantitative agreement with laboratory experiments (see, e.g., Refs. 7 and 8), the numerical experiments[13,14] performed by means of full-scale micromagnetic simulations failed to discover any self-sustained spin wave modes for bias currents below the threshold current (5) of instability of a linear "Slonczewski" mode. One possible explanation of this fact is that the instability of the bullet mode[12] is *subcritical*, i.e. requires a *finite* level of initial magnetization fluctuations to manifest itself. In laboratory experiments the necessary finite level of fluctuations could be caused by the thermal noise, or the influence of the Oersted magnetic field, or/and by any other small interaction, neglected in the micromagnetic model.

To make the excitation of subcritical modes[12, 15] possible in our simulations, even in the case when the thermal noise is ignored, we studied spin wave dynamics with progressively *increasing* and *decreasing* bias current. In this case one expects excitation of linearly unstable modes at the increasing branch of the current variation, while the excitation of nonlinear subcritical modes can be achieved when the current is progressively decreased starting from a large current value corresponding to a strongly nonlinear regime of spin wave excitation.

**NUMERICAL TECHNIQUE**

In our numerical simulations we used 3D Finite-Difference Time-Domain (FD-TD) micromagnetic code[17-20]. The dimensions of our computational region have been set as $L \times L \times d = 800$ nm $\times$ 800 nm $\times$ 5 nm and in calculations we used 3D mesh of 4 nm $\times$ 4 nm $\times$ 5 nm discretization cells. It was also necessary to specify the boundary conditions that adequately describe the experimental conditions in the current-driven nano-contact.

Because the dimensions of the computational region are substantially smaller than the physical size of the multilayer where the nano-contact was made in laboratory experiments[7-9] and relatively low magnetic damping ($\alpha_G = 10^{-2}$) of the FL, we had to impose absorbing boundary conditions at the edges of the computational region to prevent the reflections of the propagating spin wave modes from these boundaries that might occur otherwise. If the absorbing boundary conditions are not imposed, the interference between the waves propagating outwards and the ones reflected from the computational boundaries occurs in numerical simulations, and this purely computational artifact can lead to the substantial distortions in the computed picture of the phenomenon and can seriously affect the computed values of the threshold currents for linear propagating spin waves[13,14,17-22].

The problem of finding the exact analytical formulation of the perfectly absorbing conditions for spin waves at the edges of the computational region has not been solved so far. The attempts to find such conditions in numerical simulations have been undertaken[13,14,17,18,21] by assuming that magnetic dissipation in the magnetic medium of the FL increases near the borders of the computational region according to a certain chosen empirical dependence on coordinates. Here we are using a similar technique, assuming that in the middle of the region of computation ($0 < r < R^*$) the dissipation might be nonlinear[16], but is independent of the radial coordinate $r$, while close to the region boundary ($R^* < r < L/2$) the dissipation is linearly increasing with coordinate and has the spatial rate $c$:

$$\alpha(\xi,r) = \begin{cases} \alpha_G(1+q_1\xi), & \text{if } r < R^* \\ \alpha_G(1+c(r-R^*))(1+q_1\xi), & \text{if } r > R^* \end{cases}. \qquad (10)$$

The parameters of the dissipation function (10) $R^* = L/2 - 40$ nm and $c = 100/(L/2 - R^*)$ were chosen empirically to minimize the reflection of the propagating wave in a numerical experiment.

For the geometry under investigation and the parameters of our simulation we have numerically checked that the lowest reflection coefficient for propagating waves is achieved when the spatial region of damping increase near the computational region boundary involves about 10 grid cells and the final damping value is approximately two orders of magnitude larger than the Gilbert damping constant $\alpha_G$ in the middle of the computational region (which corresponds to a damping value of the order of unity). The further proof that our choice of parameters of the dissipation function (10) is reasonable comes from the fact that the threshold of excitation of a linear spin wave mode numerically calculated using the dissipation function (10) does not differ from the analogous threshold analytically calculated using Eq. (5) by more than 10 %. Using a similar criterion, we have also numerically verified that the computational region having the size 800 nm × 800 nm, which we have chosen for our simulations, is sufficiently large to give reasonable quantitative values for all the calculated quantities.

In should be noted that reflections at the boundary of the computational region can also take place because of the inhomogeneous profile of the static internal magnetic field near these boundaries. To overcome this problem one usually uses either periodic boundary conditions[13,14,23,24] or open boundary conditions[17]. For the geometry of our simulations we used a different approach.

Using the fact that our computational area is much larger than the typical wavelengths of the excited spin wave modes and expecting that the magnetization distributions calculated in our simulations would be reasonably smooth, we assumed that the magnetization far away from the nano-contact area is aligned along the direction ($\mathbf{e_z}$) of the external bias magnetic field. Thus, we assumed that at the actual boundaries of the computational region the variable magnetization is fixed and is parallel to the direction of the bias magnetic field ($z$-axis):

$$\mathbf{M}|_{\text{boundaries}} = M_0 \mathbf{e}_z. \qquad (11)$$

Outside the gridded region, the magnetization is also constrained to lie along the bias field direction. The magnetostatic charges appearing at the ends of the calculation region were consequently discarded.[25,26]

It has been checked numerically that the above described pinned boundary conditions, acting on both exchange and magnetostatic fields, worked sufficiently well, i.e. a reasonably flat profile of the total effective field has been obtained in the vicinity of the computational boundaries.

## RESULTS AND DISCUSSION

In our numerical simulations we started from the initial equilibrium state $\mathbf{M} = M_0\mathbf{e}_z$, and progressively increased the value of the applied bias current (taken with the proper sign corresponding to the case when electrons flow from FL to PL[3]). We found that at the value $I_{th}^L = 11$ mA (which constitutes our numerical threshold current for the excitation of the linear Slonczewski-like mode[3]) the initial uniform magnetization state loses its stability, and the system reaches the limit cycle representing the microwave generation. This threshold value is the same for both models of dissipation (with $q_1 = 3$ and $q_1 = 0$), and is quite close to the theoretical value $I_{th}^L = 11.5$ mA of the threshold current of the linear Slonczewski's mode calculated using Eq. (5).

As it can be seen in Figs. 2 (a) and (c), the frequency of the Slonczewski-like propagating spin-wave mode excited at the threshold is above the FMR frequency $f_{FMR} = \omega_{FMR}/2\pi = 22.5$ GHz and exhibits the expected red-shift when the magnitude of the bias current is increased. This result agrees qualitatively with the results obtained in the previous numerical simulations[13,14]. The comparison of the frequency and wave vector of the excited spin wave mode obtained in our current numerical simulations ($f_{num} = 27.2$ GHz, $k_{num} = 6.04 \times 10^7$ m$^{-1}$) with the corresponding predictions of the analytical theory[3] ($f_{lin} = 27.5$ GHz, $k_{lin} = 1.2/R_c = 6.00 \times 10^7$ m$^{-1}$) shows a satisfactory agreement between them, which proves the linear and propagating nature of the observed mode. We note that a similar agreement between the analytic theory[3] and numerical results for the linear propagating spin wave mode excited in a *perpendicularly* magnetized current-driven magnetic nano-contact has been demonstrated for in our previous work[17].

If the bias current is further increased in our simulation, an abrupt downward jump in the frequency of the excited mode is observed in both dissipation models (see Fig. 2 (a) and (c)), even though the range of existence of the linear (high-frequency) mode is larger in the case of nonlinear damping[16]. In both cases the new modes appearing after the jump have the frequency which is below the FMR frequency of the free magnetic layer, similar to the results of simulation performed for large bias currents in Refs. 13 and 14.

A similar mode-switching behavior can be seen from the numerical results obtained for the averaged precession angles of the corresponding modes (see Figs. 2 (b) and (d)). We defined the averaged precession angle $\varphi$ in a particular spin wave mode as the time-average value of the angle between the *z*-axis and the magnetization vector **M** averaged over the nano-contact area. Just above the threshold of excitation of a linear spin wave mode ($I > 11$ mA), the magnetization precesses around the direction of the in-plane external bias magnetic field (the average of the *x* and *y* magnetization components over the precession period is zero). The precession angle increases with increasing current, but remains smaller than 90 degrees. With the increase of the bias current, simultaneously with the downward jump of the excited mode frequency, the precession angle undergoes an analogous upward jump to the values larger than 90 degrees (see Fig. 2 (b) and (d)). These large values of the precession angle correspond to the precession of the magnetization vector around the direction that is antiparallel to the external bias magnetic field. This effectively means a local reversal of the average magnetization vector in the area beneath the contact. The precession angles corresponding to the both excited spin wave modes (low-amplitude linear mode and high-amplitude nonlinear mode) are shown schematically in the inset in the Fig. 2 (d).

As it was mentioned above, the high-amplitude low-frequency mode appearing suddenly at large bias currents (larger than the threshold for the excitation of the linear "Slonczewski" mode) was observed previously in numerical simulations[13,14]. However, it was not clear from Refs. 13 and 14 whether this low-frequency mode is identical to the analytically predicted "bullet" mode[12] or represents another more complicated type of high-amplitude nonlinear spin wave excitation.

To check the nature of this high-amplitude low-frequency mode, we performed numerical simulations with *decreasing* bias current. Starting from the stationary dynamic magnetization configuration that exists at sufficiently large bias current $I = 2I_{th}^L$, we progressively *decreased* the bias current value until the inverse transition from the precessional dynamic state (limit cycle) to the initial static equilibrium magnetization state (fixed point) took place. The results of simulations with decreasing bias current are shown in Fig. 2 by the branches denoted by dashed arrows. It is clear from Fig. 2 that the decrease of the bias current leads to the hysteretic behavior of both the generated frequency and the precession angle, and that the nonlinear mode having frequency below the FMR frequency of the FL continues to exist at the bias current values that are substantially lower that the threshold of excitation of the linear "Slonczewski" mode. This behavior is qualitatively the same for both dissipation models.

Thus, in our numerical simulations we were able to demonstrate that the nonlinear spin wave mode can exist in an in-plane magnetized magnetic nano-contact at such low values of the bias current, at which the linear propagating "Slonczewski" mode can not be supported. This means that, with a very high probability, the nonlinear low-frequency mode observed in our simulations is the self-localized "bullet" mode which has been predicted analytically in Ref. 12, but was not found in the previous numerical simulations [13-14]. We also believe that this localized "bullet" mode has been observed in the laboratory experiments [7-9].

It also follows from our numerical results that in the deterministic (without thermal noise) numerical simulations the spin wave "bullet" mode can only be excited by a hysteretic procedure when the bias current is first increased to a substantial supercritical value and then is gradually decreased.

In agreement with the analytical prediction[12] our simulations demonstrated that the lowest values of the bias current at which the "bullet" mode can exist ($I_{th}^{nonlin}(q_1 = 3) = 5.2$ mA in the nonlinear model of dissipation and $I_{th}^{nonlin}(q_1 = 0) = 2.0$ mA in the standard Gilbert dissipation model) are considerably lower than the threshold of excitation of the linear propagating mode that is the same for the both dissipation models and equal to $I_{th}^{L} = 11$ mA. This means that in the real experiment where thermal noise is always present the threshold of the "bullet" mode excitation will be substantially lower than the threshold of excitation of a linear propagating mode, and the "bullet" mode would be excited first when the bias current is increased. We note that the corresponding threshold currents for the excitation of the "bullet" mode calculated using the analytic formalism of Ref. 12 are $I_{th}^{nonlin}(q_1 = 3) = 6.9$ mA and $I_{th}^{nonlin}(q_1 = 0) = 1.4$ mA, so they are in reasonably good agreement with the above presented values found in our numerical simulations (see Fig. 2).

An additional property of the excited spin wave modes that can be successfully used for their identification is the spatial localization, which significantly differs for linear propagating mode (see Eq. (7)) and nonlinear self-localized "bullet" mode (Eq. (8)). The dependence of the squared amplitude $A^2$ on the distance $r$ (taken along the $z$ axis in Fig. 1) is shown in Fig. 3 for both linear and "bullet" modes. The curves in Fig. 3 were calculated for the bias current $I = 12$ mA at which both modes exist simultaneously (see Fig. 2). It is clear from Fig. 3 that the "bullet" mode is exponentially localized and at a distance $r \approx 4R_c$ from the center of the nano-contact the bullet amplitude is three orders of magnitude lower than its value at the contact center. In comparison, the amplitude of the linear propagating spin wave mode at the same distance is two orders of magnitude larger than the "bullet" amplitude.

The analysis of the numerical data on the spatial localization of the excited spin wave modes allows us to confirm the analytical conclusion[12] of the evanescent character of the "bullet" mode. In Fig. 4 we show numerically calculated profiles of the linear (dashed line) and "bullet" (solid line) modes in logarithmic scale and, for comparison, the analytical profiles (dash-dotted lines) of the linear mode calculated from Eq. (7) (Fig. 4 (a)) and of the "bullet" mode calculated from Eq. (8) (Fig. 4 (b)). It is clear, that numerical simulations are in reasonably good agreement with the predictions of the analytical model[12], which predicts that the bullet mode is a strongly localized evanescent mode. The weak oscillations of the amplitude of the linear propagating mode observed in our numerical simulations Fig. 4 are, most probably, related to the fact that the boundary conditions chosen in our simulations at the edges of spatial region of computation were not ideally absorbing, and resulted in the weak reflection of the linear propagating mode.

To further prove the evanescent character of the high-amplitude "bullet" mode we determined (using Eq. (8)) the modulus of the "bullet" wave number $|k_B|$ from the spatial profiles of the "bullet" mode numerically calculated for different values of the bias current using the fitting procedure similar to the one shown in Fig. 4 (a). Then, we plotted in Fig. 5 the "bullet" mode frequencies numerically calculated for two different dissipation models (see Fig. 2) against the above determined values of $|k_B|$ corresponding to the same values of the bias current. The results of this calculation are shown by solid squares (nonlinear dissipation with $q_1 = 3$) and open circles (Gilbert dissipation with $q_1 = 0$) in Fig. 5. In the same figure, for comparison, we show by a solid line the analytical dependence Eq. (9) of the bullet frequency $f_B$ on the modulus of the "bullet" wave number $|k_B|$. It is clear that in the whole range of calculated "bullet" frequencies the spatial localization of the "bullet" mode follows the formula (8), where the modulus of the "bullet" wave number is very close to its "evanescent" value $|k_B| = \sqrt{\omega_{FMR} - \omega_B / D}$.

Thus, it has been numerically proven that the high-amplitude spin wave "bullet" in an in-plane magnetized nano-contact having its frequency nonlinearly shifted below the FMR frequency $\omega_{FMR}$ of the FL has, indeed, evanescent character, as it was predicted in the analytical calculation[12]. We also note, that the conclusion about the evanescent nature of the "bullet" mode does not depend significantly on the dissipation model used (nonlinear or Gilbert).

**CONCLUSION**

In conclusion, using full-scale micromagnetic simulations, we have numerically proven that a current-driven in-plane magnetized magnetic nano-contact can support at least *two different types* of microwave spin wave modes: quasi-linear propagating "Slonczewski" mode[3] and the subcritically-unstable[15] self-localized nonlinear spin wave "bullet" mode[12]. We have shown that the "bullet" mode, having very large precession angles exceeding 90 degrees, can exist at the bias currents that are substantially lower than the threshold of excitation of the linear "Slonczewski mode (see Fig. 2) and, therefore, in real finite-temperature laboratory experiments[7-9], where the thermal noise is present, the "bullet" mode is the mode that is be excited first when the bias current is increased. In our zero-temperature numerical simulation, where the influence of the thermal noise is excluded, the "bullet" mode can be excited only if we reduce bias current starting from the large supercritical values of it that significantly exceed the linear spin wave mode threshold.

We have also proven that the high-amplitude "bullet" mode is an *evanescent* mode, whose spatial localization is directly related to the difference between the "bullet" frequency $\omega_B$ shifted down by the nonlinearity ("red" nonlinear frequency shift, see e.g. Eq. (6) in Ref. 12) and the FMR frequency of the "free" layer $\omega_{FMR}$.

Thus, our numerical simulations confirmed all the qualitative conclusions made in Ref. 12 about the nonlinearly-localized and evanescent nature of the spin wave "bullet" mode excited by spin-polarized current in an in-plane magnetic nano-contact.

**ACKNOWLEDGEMENTS**

The authors gratefully acknowledge discussions with Luis Lopez-Diaz.

This work was supported in part by the MURI grant W911NF-04-1-0247 from the Department of Defense of the USA, by the contract No.W56HZV-07-P-L612 from the U.S. Army TARDEC,

RDECOM, by the grant ECCS-0653901 from the National Science Foundation of the USA, and by the Oakland University Foundation.

[*]email address: [consolo@ingegneria.unime.it](mailto:consolo@ingegneria.unime.it)

# FIGURE CAPTIONS

**Fig. 1.** (Color online) Sketch of the point-contact device structure with the coordinate system used in our simulations.

**Fig. 2.** (Color online) Dependence of the precession frequency $f$ (panels (a) and (c)) and averaged precession angle $\varphi$ (panels (b) and (d)) in the excited spin wave modes on the applied bias current $I$. Panels (a) and (b) correspond to the non-linear damping model with $q_1 = 3$, while panels (c) and (d) correspond to the standard Gilbert damping ($q_1 = 0$). Arrows indicate the directions of current variation. The lines corresponding to different modes are indicated by the mode name (linear or "bullet"). The inset in (d) shows the amplitudes of the precession angle for the linear and "bullet" modes. The horizontal dashed lines in (a) and (c) show the FMR frequency.

**Fig. 3.** (Color online) Main panel: Dependence of the numerically calculated normalized squared amplitude $<A^2>/<A^2>_{max}$ (or mode power) for the "bullet" mode (solid line) and for the linear propagating mode (dashed line) on the distance $r$ from the nano-contact center. The dotted vertical line indicates the position of the nano-contact radius ($r = R_c$). Insets show the dependence of the mode power on the coordinates in the $y$-$z$ plane (see Fig. 1) for both "bullet" (a) and linear propagating (b) modes. All the graphs correspond to the bias current $I = 12$ mA when the "bullet" mode and linear spin wave mode can exist simultaneously (see Fig. 2).

**Fig. 4.** (Color online) Numerically calculated spatial profiles $A^2(r)$ of the "bullet" mode (solid line) and linear propagating mode (dashed line) shown in logarithmic scale. Dash-dotted lines show the analytical profiles of the linear (a) and bullet (b) modes calculated from Eq. (7) and Eq. (8), respectively.

**Fig. 5.** (Color online) Dependence of the frequency $f_B = W_B/2P$ of the "bullet" mode on the absolute value $|k_B|$ of the imaginary "bullet" wave number calculated from the numerical "bullet" profiles using Eq. (8) (like in Fig. 4 (b)): symbols – frequencies calculated numerically (see Fig. 2) for the nonlinear damping model with $q_1 = 3$ (solid squares) and for the standard Gilbert damping $q_1 = 0$ (open circles); solid line – "bullet" frequency analytically calculated from Eq. (9).

**Fig. 1.**

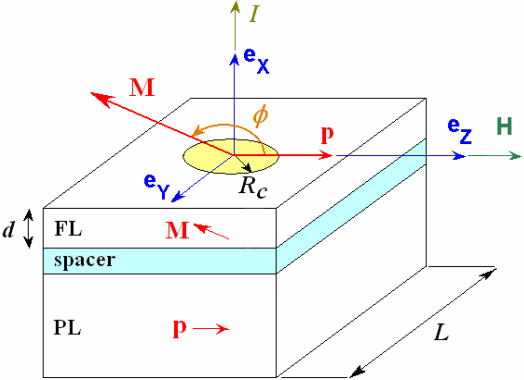

**Fig. 2.**

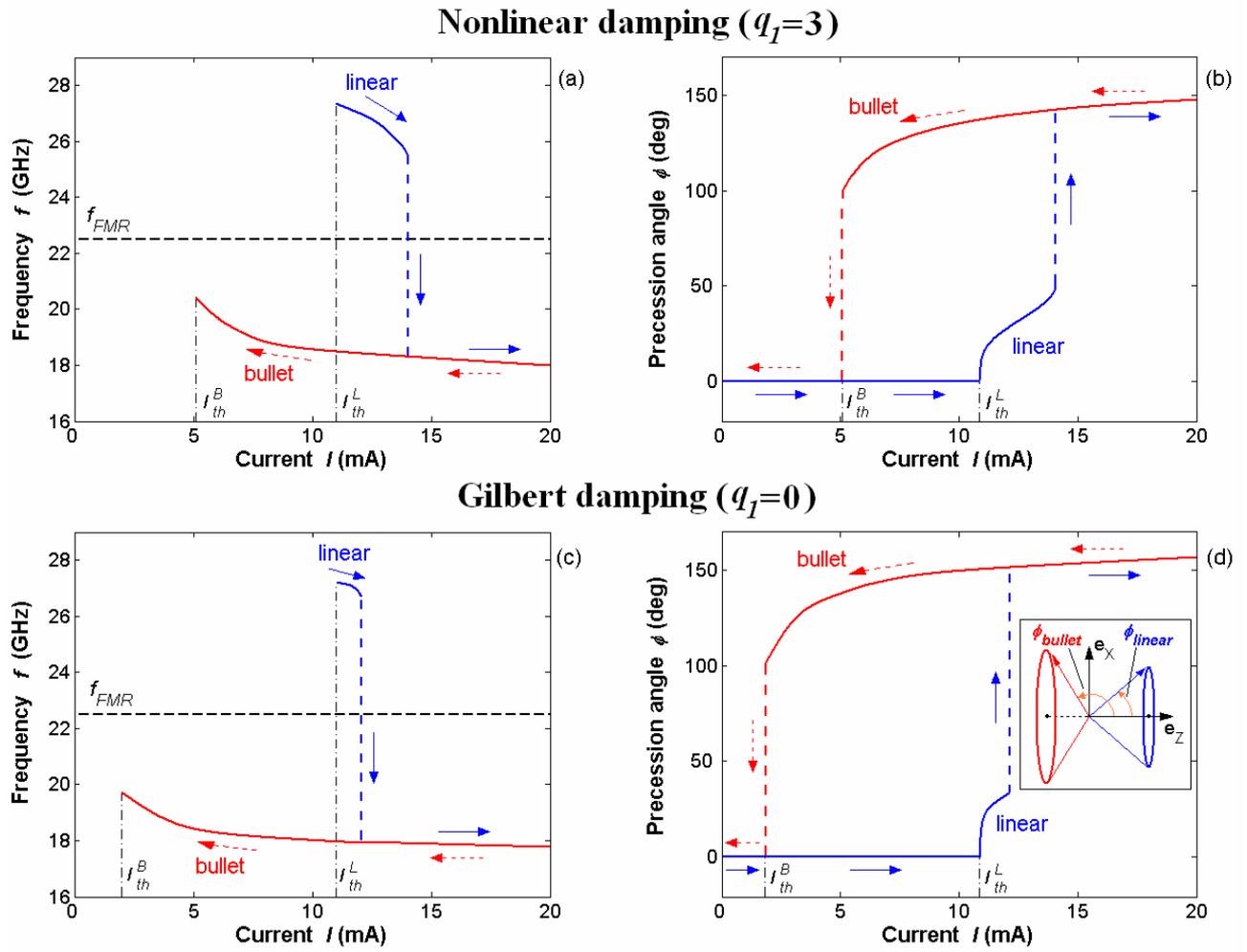

**Fig. 3.**

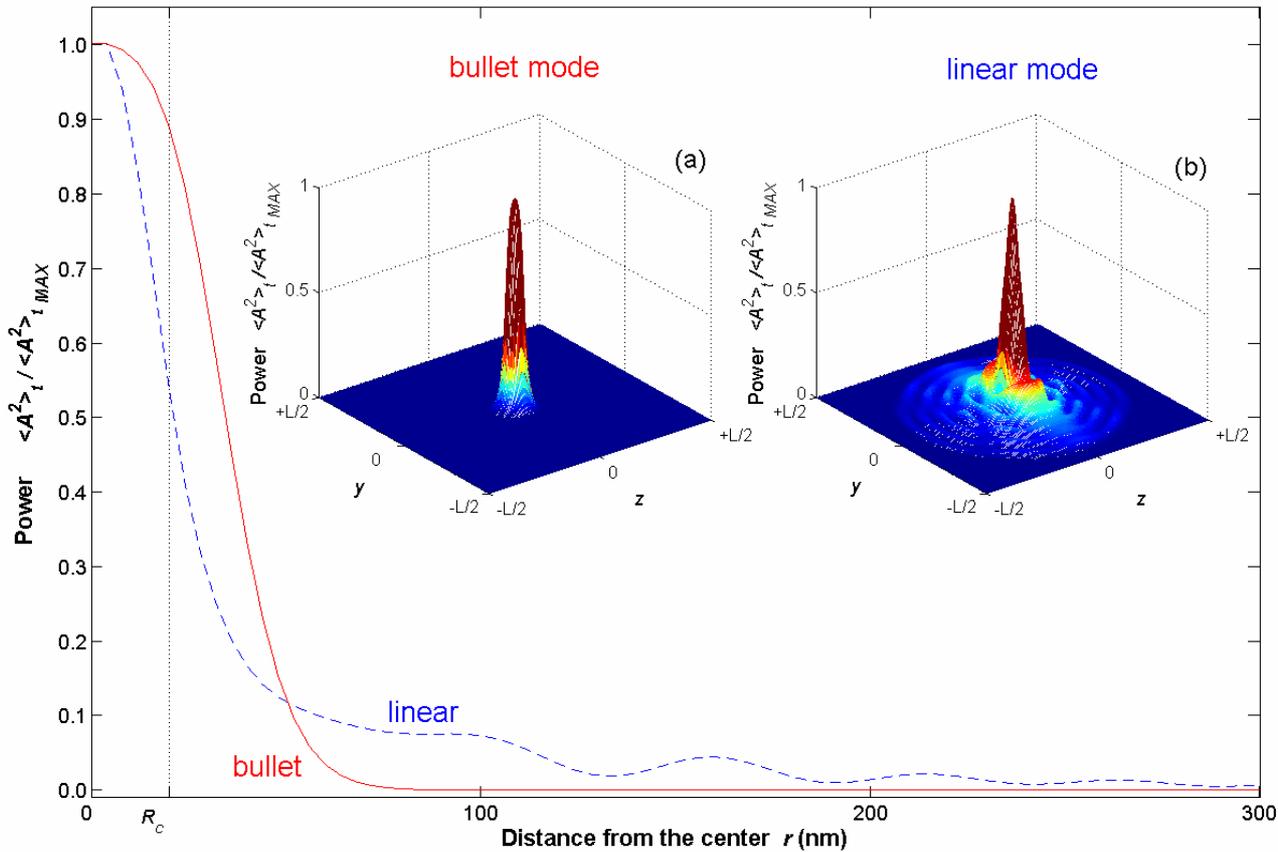

**Fig. 4.**

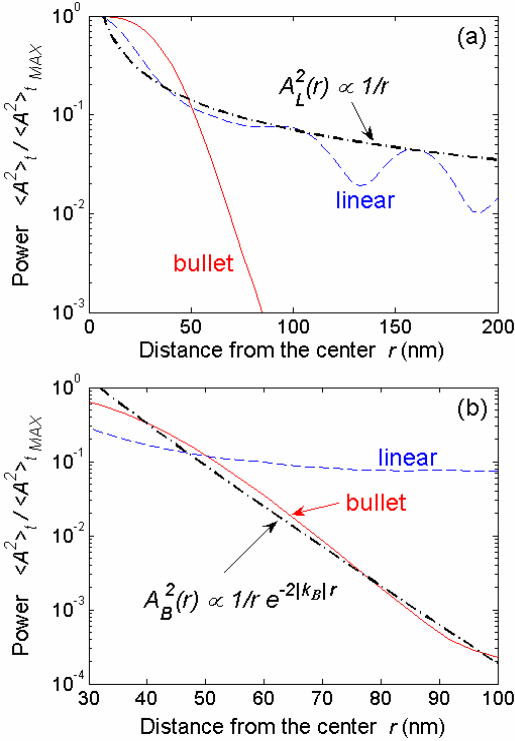

**Fig. 5.**

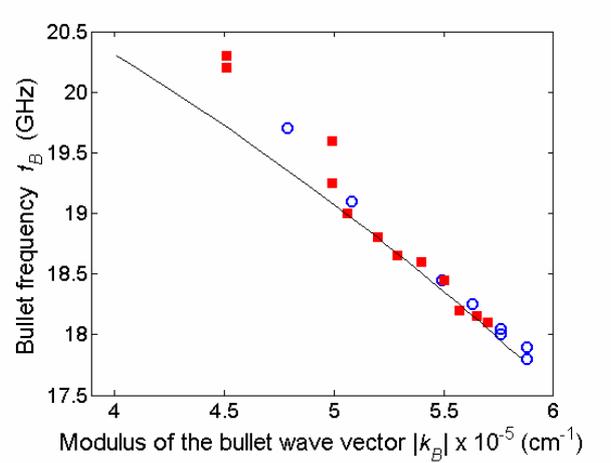